\documentclass[twocolumn,unsortedaddress,superscriptaddress]{revtex4-2}

\usepackage[english]{babel}

\addto\captionsenglish{}
\usepackage{graphicx}

\usepackage{epsfig,color}
\usepackage{amsmath,amssymb,calc}
\usepackage{amsthm}
\usepackage{enumerate}
\usepackage{hyperref}
\usepackage{amsxtra,amsfonts,bm,txfonts}
\usepackage{blkarray}
\usepackage{adjustbox}
\usepackage{braket}
\usepackage{adjustbox}
\usepackage{xcolor}
\usepackage{siunitx}
\usepackage[normalem]{ulem}

\usepackage{tikz,ifthen}
\usetikzlibrary{ positioning, quotes,math}

\definecolor{niceblu}{HTML}{ 1f618d }  
\definecolor{nicegreen}{HTML}{ 28b463 }
\definecolor{nicered}{HTML}{   a93226   }
\definecolor{niceorange}{HTML}{   d35400   }
\definecolor{nicepurp}{HTML}{   7d3c98   }

\newcommand{\fulleqref}[1]{Eq. \eqref{#1}} 
\newcommand{\figref}[1]{Fig. \ref{#1}} 
\newcommand{\fullcite}[1]{Ref. \cite{#1}} 
\newcommand{\paren}[1]{\left( {#1} \right)} 
\newcommand{\parsq}[1]{\left[ {#1} \right]} 
\newcommand{\abs}[1]{\left| {#1} \right|} 

\newcommand{\op}[1]{\hat{#1}} 
\newcommand{\uvc}[1]{#1} 

\begin{document}

\title{Truncation-Free Quantum Simulation of Pure-Gauge Compact QED Using Josephson Arrays}

\date{\today}

\author{Guy Pardo}
\affiliation{Racah Institute of Physics, The Hebrew University of Jerusalem, Jerusalem 91904, Givat Ram, Israel.}

\author{Julian Bender}
\affiliation{Center for Theoretical Physics, Massachusetts Institute of Technology, Cambridge, MA 02139, USA}
\affiliation{Max-Planck-Institute of Quantum Optics, Hans-Kopfermann-Str. 1, 85748 Garching, Germany}

\author{Nadav Katz}
\affiliation{Racah Institute of Physics, The Hebrew University of Jerusalem, Jerusalem 91904, Givat Ram, Israel.}

\author{Erez Zohar}
\affiliation{Racah Institute of Physics, The Hebrew University of Jerusalem, Jerusalem 91904, Givat Ram, Israel.}

\begin{abstract}
    Quantum simulation is one of the methods that have been proposed and used in practice to bypass computational challenges in the investigation of lattice gauge theories. 
    While most of the proposals rely on truncating the infinite dimensional Hilbert spaces that these models feature, we propose a truncation-free method based on the exact analogy between the local Hilbert space of lattice QED and that of a Josephson junction. 
    We provide several proposals, mostly semi-analog, arranged according to experimental difficulty. 
    Our method can simulate a quasi-2D system of up to $2\times N$ plaquettes, and we present an approximate method that can  simulate the fully-2D theory, but is more demanding experimentally and not immediately feasible. 
    This sets the ground for analog quantum simulation of lattice gauge theories with superconducting circuits, in a completely Hilbert space truncation-free procedure, for continuous gauge groups. 
\end{abstract}

\maketitle

\section{Introduction} \label{sec:intro}
Gauge theories are a family of models that describe the interactions between fundamental particles, and form the basic building blocks of the standard model of particle physics \cite{peskin_introduction_1995, fradkin_field_2013}. 
Since some of them include interesting non-perturbative regimes (specifically quantum chromodynamics at low energies), analytical methods struggle to accurately describe important phenomena such as the confinement of quarks into hadrons \cite{wilson_confinement_1974}.
Studying discrete formulations of these models, known as lattice gauge theories (LGTs) \cite{wilson_confinement_1974, kogut_hamiltonian_1975, kogut_introduction_1979}, has been one of the most promising research directions in answering these open questions via a continuum limit.
The lattice models themselves are ubiquitous in condensed matter physics (e.g. in high-$T_c$ superconductors), where they often emerge as effective descriptions \cite{Wen2007, sachdev_2016_emergent}; and they also feature in quantum information theory due to analogies to quantum error correction \cite{kitaev_fault-tolerant_2003}. 

Applying numerical Monte-Carlo methods to LGTs has enabled the calculation of important previously inaccessible static properties such as the hadronic spectrum (see the review \cite{aoki_flag_2020}).
However these methods struggle with real-time dynamics (being based on Euclidean time) and suffer from the so-called sign problem, which severely limits their efficiency in important cases (fermionic models with finite chemical potential) \cite{troyer_computational_2005}.
A potential alternative method that has gained a significant amount of attention in the past decade is quantum simulation (QS) - the mapping of a model of interest to a highly controllable experimental quantum device.

QS of LGTs has been proposed in various different approaches (see e.g. the reviews \cite{wiese_ultracold_2013, zohar_quantum_2016, wiese_towards_2014, dalmonte_lattice_2016, banuls_review_2020, banuls_simulating_2020, klco_standard_2022, zohar_quantum_2022, aidelsburger_cold_2022, bauer_quantum_hep_2023, Bauer_quantum_fundamental_2023, halimeh_2023_cold}), and many of them have been implemented experimentally (e.g. \cite{martinez_real-time_2016, Bernien_2017_probing, kokail_self_2019, schweizer_floquet_2019, mil_scalable_2020, yang_observation_2020, zhou_thermalization_2021, semeghini_probing_2021, Riechert_Engineering_2022, su_observation_2023, zhang_2023_observation, Surace_ab_initio_2023, Nguyen_digital_2022, Mueller_quantum_2023, davoudi_2024_scattering}) using ultra-cold atoms in optical lattices, trapped ions, and Rydberg atoms.
Superconducting circuits have also been used in many implementations of digital QS on a superconducting qubit processor (e.g. \cite{marcos_superconducting_2013, marcos_two-dimensional_2014, mezzacapo_non-abelian_2015, atas_su_2021, Atas_Simulating_2023, mildenberger_probing_2022, Pardo_2023_resource, Farrell_preparation1_2023, Farrell_preparation2_2023, farrell_2024_scalable, Farrell_2024_quantum, angelides_2024_first, Rahman_SU(2)_2021, Rahman_self_2022, Mendicelli_2023_real, Klco_SU(2)_2020, Ciavarella_Trailhead_2021, Ciavarella_Preparation_2022, Ciavarella_quantum_2023, ciavarella_2024_quantum, Charles_simulating_2024}), but not for analog or hybrid QS, based on a direct analogy between the simulator and the model.
All of the proposals for continuous gauge groups rely on a truncation of the (infinite dimensional) local Hilbert space that is associated with the gauge field \cite{zohar_quantum_2021, haase_resource_2021, bauer_efficient_2023, grabowska_2024_fully, fontana_2024_efficient}. 
While some truncation schemes have been shown to reproduce the full gauge theory in the continuum limit (for example - \cite{brower_qcd_1999}), most of them create some error in the QS (all of them if one is interested in the lattice model), and a simulation that is carried out in the full Hilbert space would therefore be advantageous.

We propose a truncation-free QS scheme for a pure-gauge U(1) LGT, the theory whose continuum limit is quantum electrodynamics (QED), in the absence of charges \cite{wilson_confinement_1974,kogut_introduction_1979}.
The proposal is based on the exact analogy between the local gauge field Hilbert space, and the Hilbert space of a Josephson junction (JJ): a standard superconducting circuit element \cite{Josephson_1962}. 
The key insight is that since the local Hilbert space is completely equivalent, designing a circuit with many junctions arranged in a particular array with the correct couplings can potentially be a good analog quantum simulator for the LGT which does not rely on any truncation. In particular for the coupling regime close to the continuum limit, where it is difficult to find suitable truncation schemes, we show that our proposal can be naturally implemented as it corresponds to the transmon regime in superconducting circuits.

A relation between QED in free space and superconductivity, and specifically the JJ, is not a new idea \cite{polyakov_quark_1977,HOSOTANI1977_compact}, and has been used, for example, to study finite temperature phase transitions in JJs \cite{Vakif_2001_phase}.
An important distinction is that while \fullcite{HOSOTANI1977_compact} derived a duality transformation from three-dimensional QED to an extended-element model for a single JJ, we  instead utilize the direct equivalence between a lumped-element JJ and the local Hilbert space in the lattice version of QED.
The two analogies are different but may very well be fundamentally related, especially considering that in order to go beyond the Hilbert-space equivalence and implement the required interactions, we ended up using a (different) duality transformation as an intermediate step (see section \ref{sec:dual}).

Arrays of JJs provide a fertile ground for the study of different kinds of physics. They are used, in different forms, for the study of quantum phase transitions \cite{Fazio2001_quantum, Mukhopadhyay2023_Superconductivity}, for quantum amplification \cite{Castellanos-Beltran2008-mc}, and as analog simulators of black-hole physics, with possible implications for quantum gravity \cite{Chen_1991_shock_wave,Nation_2009_Hawking,Katayama_2020_hawking}.
 If we adopt the most general definition of the term "array" -  a circuit with many junctions - then some superconducting qubit processors also qualify, and in fact these are the most similar to the type of JJ arrays that we propose.

The article is organized as follows: we begin by reviewing the  pure-gauge U(1) LGT in the Hamiltonian formalism (section \ref{sec:background:U1LGT}) and the physics of JJs (section \ref{sec:background:JJ}); and observe the Hilbert space analogy between the two.
Then we describe the type of JJ arrays that our proposal is based upon (section \ref{sec:background:JJArray}). 
In section \ref{sec:dual} we introduce a known dual reformulation of the original LGT \cite{drell_quantum_1979, kaplan_gauss_2018, bender_gauge_2020}, which is more easily related to a JJ array.
This is followed by a few QS proposals, arranged according to the experimental difficulty: 
In section \ref{sec:analog} we propose a fully analog QS for a (very) small system with just two plaquettes ($2\times3$ sites).
The appeal of this proposal is that it is a very simple circuit which is already fabricated (up to some design parameters) and used in many superconducting circuits labs and companies.
We then explain why the fully analog scheme cannot be extended to larger systems, which motivates a hybrid analog-digital approach (section \ref{sec:hybrid}), based on tunable coupling capacitors \cite{Materise_2023_Tunable_capacitor}.
With this approach we are able to provide proposals for larger systems, with up to $2 \times N$ plaquettes, or $3\times \paren{N+1}$ sites, but not for the fully-2D model.
In section \ref{sec:errors} we analyze the effect of finite on/off ratio of the tunable capacitors on the quality of the QS, and finally in section \ref{sec:2D} we suggest an approximate method for treating the fully-2D case, which is more demanding experimentally.
This method will require a certain degree of technological advancement before it becomes feasible, mostly due to the on/off ratio of the tunable capacitors, but a few other relevant quantitative considerations are also discussed.

\section{Background} \label{sec:background}
\subsection{U(1) Lattice gauge theory} \label{sec:background:U1LGT}
\begin{figure}
\centering
  \includegraphics[width=0.65\columnwidth]{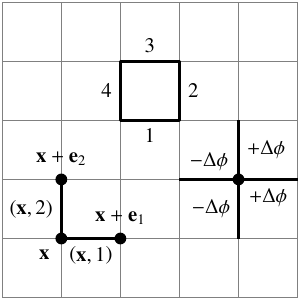}
  \caption{Geometry and conventions for two-dimensional U(1) lattice gauge theory. Each link (grey edge connecting two intersections) hosts a Hilbert space of a particle on a ring. (left) Notation for sites, links and directions of the lattice. (top) Convention for link indices within a given plaquette, used in \fulleqref{eqn:H_KS}. (right) A gauge transformation at a given site transforms the links around it, shifting the angle coordinate on outgoing links by some angle $\Delta\phi$, and on incoming links by $-\Delta\phi$. }
  \label{fig:lattice_geometry}
\end{figure}

Our model of interest is the pure-gauge U(1) LGT, defined on a two-dimensional square lattice with sites $\mathbf{x}=\paren{x_1,x_2}\in \mathbb{Z}^2$ and links $\paren{\mathbf{x},i}$ where $i=1,2$ indicates one of the two lattice directions (left-to-right or down-to-up). 
In the common convention $\paren{\mathbf{x},i}$ denotes the link that connects the site $\mathbf{x}$ and the site $\mathbf{x} + \uvc{\mathbf{e}}_i $ where $\uvc{\mathbf{e}}_i$ is a unit vector in the direction $i$ (see \figref{fig:lattice_geometry}).
With each link is associated a Hilbert space of a particle on a ring, with an angular (\emph{compact}) position operator $\op{\phi}_i\paren{\mathbf{x}}$ and conjugate angular momentum operator $\op{E}_i\paren{\mathbf{x}}$ which is often referred to as the \emph{electric field} because of its role in the continuum theory (with the lattice spacing approaching zero).
It follows from the compactness of $\op{\phi}_i\paren{\mathbf{x}}$ that $\op{E}_i\paren{\mathbf{x}}$ has an unbounded integer spectrum.

\emph{Gauge transformations} are a specific kind of local transformations that are associated with the sites.
A gauge transformation at $\mathbf{x}$ shifts the angles of the two links that come out of $\mathbf{x}$ (in the positive $\uvc{\mathbf{e}}_i$ directions) by the same angle $\Delta\phi$, and by $-\Delta\phi$ for the two links that go into $\mathbf{x}$ (from the negative $\uvc{\mathbf{e}}_i$ directions, see \figref{fig:lattice_geometry}). 
Since $\op{E}_i\paren{\mathbf{x}}$ is the generator of translations in $\op{\phi}_i\paren{\mathbf{x}}$, these transformations are generated by 

\begin{equation} \label{eqn:gauss_law_operator}
    \op{G}\paren{\mathbf{x}}=\sum_i \paren{\op{E}_i\paren{\mathbf{x}} - \op{E}_i\paren{\mathbf{x}-\uvc{\mathbf{e}}_i}},    
\end{equation}
which is the lattice divergence of the electric field. 

By assumption, the Hamiltonian is \emph{gauge-invariant}, which means that it is invariant under all gauge transformations, or equivalently that it commutes with $\op{G}\paren{\mathbf{x}}$ for all $\mathbf{x}$. 
A conventional choice is the Kogut-Susskind Hamiltonian \cite{kogut_hamiltonian_1975, kogut_introduction_1979}
\begin{equation}\label{eqn:H_KS} 
\begin{aligned} 
    \op{H}_{\text{KS}} &= -\frac{1}{g^2}\sum_\text{plaq.}\cos\paren{\op{\phi}_1 +\op{\phi}_2 -\op{\phi}_3 -\op{\phi}_4} + \frac{g^2}{2} \sum_{\mathbf{x},i}\op{E}^2_i\paren{\mathbf{x}}\\  
                    &\equiv \op{H}_B+\op{H}_E
\end{aligned}    
\end{equation}
(where $g^2$ is the \emph{coupling constant}), which is a simple but nontrivial Hamiltonian constructed to obey this condition. 
The first (\emph{magnetic}) term is a sum over plaquettes, with the indices $1$-$4$ labeling the four different links that form a given plaquette and following the convention introduced in \figref{fig:lattice_geometry}. 
In the large $g^2$ regime one can treat $\op{H}_\text{KS}$ perturbatively (the interaction is weak), and the well-studied electric-basis truncation is suitable \cite{bauer_efficient_2023}.
However in the $g^2<1$ regime, where the continuum limit is well-defined, both perturbation theory and the electric-basis truncation fail; and it is in this regime that our QS proposals can be advantageous.

Only gauge-invariant states and operators are considered as physically meaningful, and therefore the so-called \emph{physical states} $\ket{\Psi}$ are those that obey the \emph{Gauss law} constraint:
\begin{equation} \label{eqn:gauss}
    \op{G}\paren{\mathbf{x}}\ket{\Psi} = q(\mathbf{x})\ket{\Psi} \hspace{10pt} \forall \mathbf{x} ,
\end{equation}
where the \emph{static charges} $q\paren{\mathbf{x}}$ are constants of motion that split the Hilbert space into \emph{superselsction sectors}.
Each sector is characterized by a particular charge configuration, 
and the gauge-invariance of $\op{H}_\text{KS}$ ensures that the dynamics do not include transitions between different sectors. 
It is therefore always assumed that $q\paren{\mathbf{x}}$ are fixed, and it is common to choose $q\paren{\mathbf{x}}=0 \hspace{3pt} \forall \mathbf{x}$ (no static charges).  

From the point of view of QS, this redundancy of the Hilbert space is undesirable since it leads to wasting expensive quantum resources, and it requires monitoring and enforcement of the constraints. 
For this reason many redundancy-free formulations have been developed for LGTs \cite{hamer_lattice_1979, banuls_efficient_2017, martinez_real-time_2016, sala_variational_2018, zohar_eliminating_2018, zohar_removing_2019, irmejs_quantum_2022, Pardo_2023_resource, raychowdhury_loop_2020, kadam_loop_2022, Kadam_2023_loop, bauer_efficient_2023,grabowska_2024_fully, ciavarella_2024_quantum, FONTANA_2022_reformulation}, and one of them \cite{drell_quantum_1979, kaplan_gauss_2018, bender_gauge_2020} is also helpful for constructing a superconducting circuit analogy, as we will show in the following.

\subsection{The Josephson junction Hamiltonian} \label{sec:background:JJ}
The principal component in most superconducting circuit applications is the Josephson junction (JJ), which operates as a nonlinear inductor.
It consists of two superconducting electrodes separated by some kind of a \emph{weak link}, or an obstruction that is thin enough to allow Cooper-pairs to tunnel through.

As we mentioned in section \ref{sec:intro}, the Hilbert space of a JJ is completely equivalent to the link Hilbert space of the U(1) LGT as presented in section \ref{sec:background:U1LGT}.
The (gauge-invariant) phase difference $\Delta\varphi$ of the superconducting wavefunction across the junction is a compact quantum degree of freedom, and it can be related (equated modulo $2\pi$) to the \emph{reduced magnetic flux} through the junction $\phi \equiv 2\pi \Phi/\Phi_0$, where $\Phi$ is the magnetic flux and $\Phi_0 = h/2e$ is the magnetic flux quantum.
Using $\Delta \varphi$ and $\phi$ interchangeably is a common abuse of notation, which is legitimate as long as we are being careful to write down only $2\pi$-periodic functions of $\phi$.
The canonical conjugate to the reduced flux is the \emph{reduced charge} $n \equiv Q/2e$ (where $Q$ is the electrical charge), which is the excess number of Cooper-pairs on the two electrodes and takes integer values.
Thus, by identifying the reduced flux operator $\op{\phi}$ through a junction with the angular position operator $\op{\phi}_i\paren{\mathbf{x}}$ on a link, and the reduced charge operator $\op{n}$ of a junction with the electric field operator $\op{E}_i\paren{\mathbf{x}}$ of a link, the equivalence of the two Hilbert spaces is manifested. 

The dynamics of tunneling through the JJ can be understood via the following Hamiltonian \cite{devoret}
\begin{equation} \label{eqn:tunnel_ham}
    \op{H}_{\text{tunneling}} = -\frac{E_J}{2}\sum_{n=-\infty}^\infty \ket{n}\bra{n+1} + \text{h.c.},
\end{equation}
where $\ket{n}$ is the macroscopic state with an imbalance of $n$ pairs between the two electrodes (eigenstate of the reduced charge operator $\op{n}$).
The Josephson energy scale $E_J$ is proportional to the normal-state tunnel conductance and the superconducting gap, and it can also be related to the JJ critical current $I_c$ (beyond which the junction becomes resistive) via \cite{Josephson_1962}
\begin{equation} \label{eqn:I_c}
    I_c = \frac{2e}{\hbar} E_J.
\end{equation}
In the flux basis the tunneling Hamiltonian is diagonal and can be expressed as
\begin{equation} \label{eqn:tunneling_ham_flux}
    \op{H}_{\text{tunneling}} = -E_J \cos{\op{\phi}}.
\end{equation}

Any real junction should be modeled as the pure non-linear inductance element described by $\op{H}_{\text{tunneling}}$, connected in parallel to a capacitor to account for the capacitance between the two electrodes (and any other shunt capacitance that might be introduced on purpose when designing the circuit).
For this reason the Hamiltonian for a realistic Josephson junction is 
\begin{equation} \label{eqn:JJ_ham}
    \op{H}_J = 4E_C \op{n}^2 - E_J \cos{\op{\phi}},
\end{equation}
where $E_C = e^2/2C$ is the charging energy of the (total) capacitance $C$ by a single electron (and the factor of $4$ is due to the fact that $\op{n}$ is defined as the number of pairs).

If $E_J \gg E_C$ (achieved by designing a large parallel capacitor) then $\op{H}_J$ can be approximated for low energies as a weakly anharmonic oscillator with negative anharmonicity $-E_C/\hbar $ (the difference between consecutive transition frequencies).
This design is the most common implementation of a superconducting qubit (the \emph{transmon qubit}), and it utilizes the weak-but-not-insignificant anharmonicity to address only the first two levels with microwave pulses, with minimal leakage outside the computational subspace \cite{Koch_2007_charge-insensitive}.
We, however, are interested in taking advantage of the full Hilbert space and develop a quantum simulation scheme without truncation, which means that low-energy approximations are not going to be good enough for us even if we choose to operate within the transmon regime  $E_J \gg E_C$.
Therefore our starting point is \fulleqref{eqn:JJ_ham} and not the linearized transmon Hamiltonian.

In a circuit with two identical JJs connected in a loop, the two phase variables are related by the \emph{fluxoid quantization condition} \cite{Tinkham}. Therefore there is only one independent angular coordinate, that we will denote as $\phi$. 
In this case one can show that the Hamiltonian is
\begin{equation} \label{eqn:squid_ham}
    \op{H}_\text{SQUID} = 4E_C \op{n}^2 - 2E_J \abs{\cos{\paren{\pi\frac{\Phi_\text{ext}}{\Phi_0}}}}\cos{\op{\phi}},
\end{equation}
where $E_J$ is the Josephson energy of the individual junctions and $\Phi_\text{ext}$ is the applied external magnetic flux through the loop. As before $\op{n}$ is the conjugate momentum to $\op{\phi}$.
The two-junction loop (also called SQUID) behaves like a single junction with effective Josephson energy $2E_J \abs{\cos{\paren{\pi\Phi_\text{ext}/\Phi_0}}}$ that can be tuned by applying external magnetic flux.
The tunability range can be modified by using an asymmetrical SQUID with non-identical junctions. 
In the following we will always consider the single junction as an elementary building block, while keeping in mind that the Josephson energy $E_J$ can be made tunable by replacing each junction with a loop of two junctions, with energies that sum up to $E_J$ (e.g. $E_J/2$ each, or some asymmetrical combination depending on the tunability requirement).

\subsection{Capacitively coupled Josehpson junction arrays} \label{sec:background:JJArray}

\begin{figure}
    \centering
    \includegraphics[width=\columnwidth]{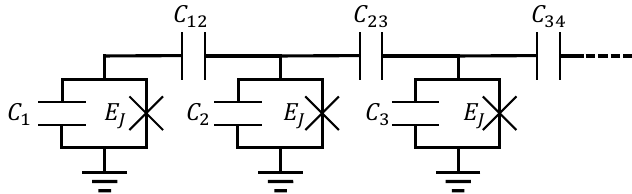}
    \caption{Circuit diagram for an example of a capacitively-coupled Josephson array. The JJs in the array have identical Josephson energies $E_J$, and the $i$th junction is shunted to the ground via its self capacitance $C_i$. The coupling capacitor between the $i$th and $j$th JJs is denoted as $C_{ij}$. The diagram shows a one-dimensional array but the formalism, summarized by \fulleqref{eqn:CCJA_ham}, is more general and allows for any two junctions to be coupled or not.  }
    \label{fig:CCJA_circuit}
\end{figure}

Our proposal is based on superconducting circuits like the one depicted in \figref{fig:CCJA_circuit}, that is, circuits with multiple junctions that are coupled to each other via some capacitors. 
We call such a circuit a \emph{capacitively-coupled Josephson array} (CCJA), because the term \emph{Josephson array} is typically used for an array of superconducting islands connected via JJs \cite{Fazio2001_quantum}. Also note that our CCJAs are different than arrays like in \fullcite{Choi_2000_Capacitively} since in our design one electrode of each junction is grounded, and only active electrodes can possibly be coupled.
In that sense our design is more similar to a superconducting qubit processor than to a metamaterial-type Josephson array; but with the important difference that we aim at involving the full Hilbert space of each JJ, rather than limit the dynamics to a truncated low-energy subspace.

We assume that all the junctions in the array have the same Josephson energy $E_J$, and the $i$th junction is shunted to the ground via a capacitor $C_i$ (its \emph{self capacitance}).
We denote the coupling capacitance between junctions $i$ and $j$ as $C_{ij}$ (see \figref{fig:CCJA_circuit}).
The Hamiltonian for a general CCJA is given by \cite{devoret}
\begin{equation} \label{eqn:CCJA_ham}
    \op{H}_\text{CCJA} = -E_J\sum_i \cos{\op{\phi}_i} + \frac{1}{2} \paren{2 e}^2 \sum_{ij} \op{n}_i \parsq{C^{-1}}_{ij} \op{n}_j,
\end{equation}
where $\parsq{C^{-1}}$ is the inverse capacitance matrix, with the capacitance matrix $\parsq{C}$ constructed according to 
\begin{equation} \label{eqn:cap_mat_def}
\begin{aligned}
    \parsq{C}_{ij} &= \parsq{C}_{ji}=  -C_{ij} \hspace{5pt} &\text{for $i\neq j$} \\
    \parsq{C}_{ii} &= C_i + \sum_{j\neq i} C_{ij}. & 
\end{aligned}  
\end{equation}
The off-diagonal elements of $\parsq{C}$ are minus the coupling capacitances, and each diagonal element $\parsq{C}_{ii}$ is the sum of all capacitances connected directly to node $i$  (including the self capacitance).
In typical implementations (e.g. for coupling of superconducting qubits) the coupling capacitors are small compared to the self capacitances.
This is useful because in this case if $\parsq{C}$ is local (e.g. includes only nearest-neighbours coupling) then $\parsq{C^{-1}}$ is also local to a good approximation.
In contrast, if the coupling capacitors are comparable to the self capacitance, then in general a local $\parsq{C}$ does not imply a local Hamiltonian (and vice versa).
This will become important in section \ref{sec:analog}.

\section{Dual formulation} \label{sec:dual}
\begin{figure}
\centering
  \includegraphics[width=0.95\columnwidth]{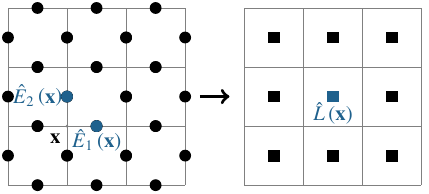}
  \caption{Illustration of the dual reformulation of the pure-gauge U(1) LGT, introduced in section \ref{sec:dual}. Originally the degrees of freedom, represented here by the electric field operator $\op{E}_i\paren{\mathbf{x}}$, are associated with the links. In the dual formulation we instead associate a loop variable $\op{L}\paren{\mathbf{x}}$ with each original plaquette or (equivalently) with sites $\mathbf{x}$ of the dual lattice}
  \label{fig:dual}
\end{figure}

Having established the Hilbert-space equivalence, the next step is to try to construct an analogy between the Hamiltonians.
The electric part $\op{H}_E$ is not too difficult since it is local, and on a single link it is already analogous to the electric part of the JJ Hamiltonian (under the identification $\op{n}\equiv \op{E}$). 
The magnetic part ${\op{H}_B}$ however is much more challenging because it is a four-body interaction and does not come up naturally in superconducting circuits. 
It seems that arranging 5 junctions in a loop (such that there are 4 degrees of freedom due to the fluxiod quantization condition) may result in the correct term under extremely careful fine-tuning of the junction parameters, but this results in many unwanted terms, and it is not clear how to scale it up to more than a single plaquette \cite{Qiu2016_four-junction}.

For this reason we use a dual reformulation of the original theory that has some advantageous properties  \cite{drell_quantum_1979, kaplan_gauss_2018, bender_gauge_2020}. 
First, the local Hilbert space is still that of a particle on a ring, and therefore still equivalent to the JJ Hilbert space.
Second, in the dual formulation the plaquette term $\op{H}_B$ becomes non-interacting, and has a form that appears naturally in superconducting circuits, while the electric term becomes a two-body interaction.
The third advantage is that the formulation is gauge redundancy-free, which means that all possible states in its Hilbert space are physical states, and no constraints are needed. 
Therefore, as opposed to quantum simulations of lattice gauge theories in the original gauge-redundant formulation, experimental errors cannot break gauge-invariance and lead to unphysical results. 

In the following we provide a brief overview of the construction of the dual formulation, for the details we refer to \fullcite{drell_quantum_1979,bender_gauge_2020}.
Since we assume no static charges $q\paren{\mathbf{x}} = 0  \hspace{5pt}\forall \mathbf{x}$, the constraint on the physical states is that the electric field is transverse (divergence-free)
\begin{equation} \label{eqn:trans_gauss}
    \sum_i \paren{\op{E}_i\paren{\mathbf{x}} - \op{E}_i\paren{\mathbf{x}-\uvc{\mathbf{e}}_i}} \ket{{\Psi}} = 0 \hspace{10pt} \forall \mathbf{x}.
\end{equation}
In order to remove this redundancy we define a new set of variables that respects the transverse nature of the field.
Since divergence-free configurations are made of loops, it makes sense to associate a \emph{loop variable} $\op{L}\paren{\mathbf{x}}$ with each plaquette of the original model, or equivalently with each site $\mathbf{x}$ of the dual lattice (see \figref{fig:dual}), such that the electric field on a link is given by the lattice-curl of $\op{L}\paren{\mathbf{x}}$:
\begin{equation}\label{eqn:loop_variable}
\begin{aligned} 
    \op{E}_1\paren{\mathbf{x}} &= {\op{L}\paren{\mathbf{x}} - \op{L}\paren{\mathbf{x} - \uvc{\mathbf{e}}_2}} \\
    \op{E}_2\paren{\mathbf{x}} &= -\paren{\op{L}\paren{\mathbf{x}} - \op{L}\paren{\mathbf{x} - \uvc{\mathbf{e}}_1}},
\end{aligned}
\end{equation}   
and is therefore transverse by construction, resulting in a redundancy-free formulation. 
\fulleqref{eqn:loop_variable} holds for open boundary conditions, which we assume from now on (for periodic boundary conditions there is a small complication because one has to consider global loops as well). 
The canonical conjugate to $\op{L}\paren{\mathbf{x}}$ is a compact variable, denoted $\op{B}\paren{\mathbf{x}}$ because it approaches the magnetic field $\mathbf{B} = \mathbf{\nabla}\times\mathbf{A}$ in the continuum limit.

Writing the transformed $\op{H}_\text{KS}$ in terms of the new variables, one arrives at the dual formulation Hamiltonian
\begin{equation} \label{eqn:H_dual}
    \op{H}_\text{dual} =  -\frac{1}{g^2} \sum_\mathbf{x} \cos{\paren{\op{B}\paren{\mathbf{x}}}}
    +\frac{g^2}{2}\sum_{\mathbf{x},i} \paren{\op{L}\paren{\mathbf{x}} - \op{L}\paren{\mathbf{x} - \uvc{\mathbf{e}}_i}}^2,
\end{equation}
in which the magnetic part is local, and the electric part is a two-body interaction between loop variables. 
If the original model is defined on an $\paren{N+1}\times \paren{N+1}$ square lattice, then it has $2N\paren{N+1}$ degrees of freedom (the number of links) and $\paren{N+1}^2 -1$ independent constraints (the constraint in one site is redundant), so $N^2$ physical degrees of freedom.
In the dual formulation there are again $N^2$ degrees of freedom (the number of plaquettes), and no constraints; which shows that the redundancy is completely removed.

The dual reformulation can also be applied in the case of static charges $q\paren{\mathbf{x}}\neq 0$ for some $\mathbf{x}$, via a unitary transformation that brings the constraint to the form \eqref{eqn:trans_gauss} in the transformed physical space.
However, this shifts the spectrum of $\op{E}\paren{\mathbf{x}}$ by a fractional (non-integer) offset \cite{bender_real_2020, drell_quantum_1979}, ruining the equivalence to the JJ Hilbert space.

Next, we introduce the sub-lattice transformation:
\begin{equation} \label{eqn:sub_lattice_trans}
    \begin{aligned}
        \op{B}\paren{\mathbf{x}}&\rightarrow \paren{-1}^{x_1+x_2} \op{B}\paren{\mathbf{x}}\\
        \op{L}\paren{\mathbf{x}}&\rightarrow\paren{-1}^{x_1+x_2}\op{L}\paren{\mathbf{x}}
    \end{aligned}
\end{equation}
to flip the phase of the odd sub-lattice plaquettes, such that the final version of our model Hamiltonian is 
\begin{equation} \label{eqn:H_U1}
    \op{H}_\text{U(1)} =  -\frac{1}{g^2} \sum_\mathbf{x} \cos{\paren{\op{B}\paren{\mathbf{x}}}}
    +\frac{g^2}{2}\sum_{\mathbf{x},i} \paren{\op{L}\paren{\mathbf{x}} + \op{L}\paren{\mathbf{x} - \uvc{\mathbf{e}}_i}}^2.
\end{equation}
As we will see, this version of the Hamiltonian, with the plus sign in the electric term, will be easier to implement with a superconducting circuit.
At this stage we use the equivalence to the JJ Hilbert space and identify $\op{B}\paren{\mathbf{x}}\equiv \op{\phi}_i$ and $\op{L}\paren{\mathbf{x}}\equiv \op{n}_i$, where $i$ indicates a specific junction in the CCJA, that we associate with a specific site $\mathbf{x}$ on the dual lattice (or a plaquette in the original lattice). Substituting into \fulleqref{eqn:H_U1} and opening the brackets in the second sum, we have
\begin{equation} \label{eqn:HU1_final}
    \op{H}_{\text{U(1)}} = -\frac{1}{g^2}\sum_{i}\cos{\op{\phi}_i}  + \frac{g^2}{2}\paren{4\sum_i \op{n}_i^2 + 2\sum_{\left< i,j\right>}\op{n}_i \op{n}_j},
\end{equation}
where $\left< i,j\right>$ denotes nearest neighbors on the dual lattice.
Note that in \fulleqref{eqn:H_U1} we have a sum over links.
Therefore when opening the brackets, each $\op{n}_i^2$ appears 4 times in the sum: bulk plaquettes participate in 4 different links, and boundary links still contribute a $\op{n}_i^2$ term of their associated plaquette.

By comparing \fulleqref{eqn:CCJA_ham} and \eqref{eqn:HU1_final}, we see that if we associate a junction with each plaquette, we have an exact analogy in the magnetic term, and the problem reduces to engineering the required capacitance matrix between the junctions to simulate the electric part. 
It is also evident that the scaling of the ratio $E_J/E_C$ should relate to the coupling constant via
\begin{equation} \label{E_J/E_C}
    \frac{E_J}{E_C} \propto \frac{1}{g^4}.
\end{equation}
This means that the regime for which QS is relevant (small coupling, see section \ref{sec:background:U1LGT}) coincides with the transmon regime of the junctions (large $E_J/E_C$, see section \ref{sec:background:JJ}).
Since the transmon is the industry's preferred design for superconducting qubits, our proposals can benefit from the accumulated experimental know-how in fabricating and manipulating JJs in this regime.

\section{Analog quantum simulation proposal for two plaquettes} \label{sec:analog}

 In this section we show how to implement the correct capacitance matrix for two plaquettes, and why it cannot be directly generalized to larger systems.
 This limitation motivates the hybrid approach described in section \ref{sec:hybrid}. For two plaquettes, the required inverse capacitance matrix obeys
 \begin{equation} \label{eqn:C_inv_analog_2_plaquettes}
    \parsq{C^{-1}}\propto \begin{pmatrix}
                4 & 1 \\
                1 & 4 
\end{pmatrix}.
\end{equation}
This can easily be achieved by designing the self capacitances of the two junctions to be 3 times the coupling capacitor $C_1 = C_2 = 3 C_{12}$, such that the Hamiltonian for this circuit, from \fulleqref{eqn:CCJA_ham} - \eqref{eqn:cap_mat_def}, is 
 \begin{equation} \label{eqn:H_CCJA_2_plaquettes}
    \op{H}_\text{CCJA} = -E_J\sum_{i=1,2}\cos{\op{\phi}_i} + \frac{1}{2} \frac{4e^2}{15 C_{12}} \paren{4 \sum_{i=1,2} \op{n}_i^2 + 2\op{n}_1 \op{n}_2 }.
\end{equation}
Taking advantage of the tunability of the Josephson energy (see section \ref{sec:background:JJ}), we can simulate different $g^2$ values in the same experiment by tuning  
 \begin{equation} \label{eqn:EJ_2_plaquettes}
    E_J = \frac{4e^2}{15 C_{12}} \frac{1}{g^4},
\end{equation}
which results in an exact analogy
 \begin{equation} \label{eqn:2_plaquettes_analogy}
    \op{H}_\text{CCJA} = \frac{4e^2}{15 C_{12}} \frac{1}{g^2} \op{H}_{\text{U(1)}}.
\end{equation}
Note that here $\op{H}_\text{CCJA}$ is the experimental Hamiltonian, which has dimensions of energy; while $\op{H}_\text{U(1)}$ is the model Hamiltonian, which is dimensionless.
The two-plaquettes proposal requires a very simple superconducting circuit - two junctions (or pairs of junctions, for tunability) with a strong capacitive coupling.
The simplicity of this specific experiment makes it attractive as a benchmark for the more general equivalence or to test different truncation schemes by comparison. 

A direct generalization for more than two plaquettes is not available, since in the general case the inverse capacitance matrix has to obey
\begin{equation} \label{eqn:analog_inv_C}
    \parsq{C^{-1}}_{ij} \propto \begin{cases}
        		4 & i=j\\
            1 & \text{$\left< i,j \right>$  are nearest neighbors}\\
            0 &\text{otherwise}.
    \end{cases}
\end{equation}
This cannot be implemented by setting static analog properties of the system, because of two related issues. 
First, the off-diagonal elements are the same order of magnitude as the diagonal ones, meaning that the coupling is strong.
As we mentioned in section \ref{sec:background:JJArray}, this implies that in order to have a local  $\parsq{C^{-1}}$ (coupling only nearest-neighbors on the dual lattice), $\parsq{C}$ has to be highly non-local, and therefore this scheme is not scalable. In other words, it requires an architecture with all-to-all physical coupling, in order to have a local coupling in the Hamiltonian.
Moreover, typically many of the non-local elements of the required $\parsq{C}$ are positive, which is not physical since it implies a negative coupling capacitance value.
For example, for a chain of three plaquettes,  with the required inverse capacitance matrix obeying
\begin{equation} \label{eqn:C_inv_analog_3_plaquettes}
    \parsq{C^{-1}}\propto \begin{pmatrix}
                4 & 1 & 0\\
                1 & 4 & 1\\
                0 & 1 & 4
\end{pmatrix},
\end{equation}
the required capacitance matrix has to be
\begin{equation} \label{eqn:C_mat_analog_3_plaquettes}
    \parsq{C}\propto \begin{pmatrix}
                15 & -4 & 1\\
                -4 & 16 & -4\\
                1 & -4 & 15
\end{pmatrix}.
\end{equation}
This is not an allowed capacitance matrix because it has positive off-diagonal elements, implying a negative capacitance value between nodes 1 and 3. 
This means that even if we ignore the scalability problem and focus on small systems, we cannot construct the required capacitance matrices for anything that is more complicated than a single pair of plaquettes. 

Faced with this problem, we considered using two known alternatives for indirect coupling between transmon qubits.
The first one is coupling through an intermediate off-resonance transmon coupler, which effectively couples the two main qubits with a controllable coupling coefficient that can be positive or negative \cite{YAN_2018_tunable, Sung_2021_Realization}. 
The other one is the effective coupling of spatially separated transmon qubits through a long waveguide.
By designing the separation length to fit the transition frequency of the qubits, one may induce an effective strong coupling with minimal energy loss into the waveguide itself \cite{Albrecht_2019_Subradiant, Lalumiere_2013_input_output}.
Both of these methods, however, assume that the transmon levels are almost evenly spaced, and that all relevant transition frequencies can be taken as equal. 
This is true for transmon qubits because they have a small anharmonicity, and in typical applications only the lower levels are excited.
While we may assume the former in some cases (specifically in the   $g^2 \ll 1$ regime), we do not assume the latter since we want to take advantage of the exact Hilbert space analogy. 
Assuming that only low energy states are participating in the dynamics would be essentially a truncation of the Hilbert space, which we want to avoid.

\section{Hybrid quantum simulation proposals} \label{sec:hybrid}
Since all we can do by direct analog design is to implement the interaction between two adjacent plaquettes, it is natural to think of a hybrid analog-digital approach, in which complicated interactions can be constructed out of simple "primitive"  ones.
The idea is to split the model Hamiltonian $\op{H}_\text{U(1)}$ into a few different parts, such that (1) we can implement an exact analogy for each part and (2) we are able to turn on and off the different parts in a controllable way during an experiment. 
Assuming the control is fast enough, one can alternate between the different parts for short periods of time, implementing a Trotter decomposition (a well-studied controlled approximation) of the original Hamiltonian \cite{trotter_on_1959, Suzuki1985a}.

To obey requirement (1), we have to split $\op{H}_\text{U(1)}$ such that each part includes only pairwise interactions between plaquettes.
For requirement (2) we need a way to change capacitance values in real time.
This can be achieved using tunable capacitors such as the ones suggested by \fullcite{Materise_2023_Tunable_capacitor}, in which an InAs/InGaAs heterostructure is fabricated beneath the superconducting capacitor plates. 
The electron concentration in the semiconductor two-dimensional electron gas is controlled via voltage gating, resulting in a tunable effective distance between the plates.
\fullcite{Materise_2023_Tunable_capacitor} predicts an on/off ratio of about $40$ in the capacitance values in this design.
In section \ref{sec:errors} we estimate the errors that the finite on/off ratio introduce to our quantum simulation, but in this section we assume an ideal tunable capacitor that can be turned off completely (to zero capacitance).

The ability to turn-off all the interactions between the junctions makes state-preparation and measurement straightforward.
With the interactions turned off,  standard superconducting qubit  control can be employed for state preparation, and standard readout techniques can be used to measure the populations of the non-interacting eigenstates \cite{Nakamura_coherent_1999,Koch_2007_charge-insensitive, Liu_performing_2023}.
Since the wavefunctions of these eigenstates are known (these are Mathieu functions \cite{Wilkinson_approximate_2018}), any local expectation value can be measured by repeating the experiment enough times to obtain population statistics.
This type of measurement introduces truncation, since the standard readout techniques are limited in the sense that they cannot distinguish between infinitely many eigenstates.
However, the truncation only happens at the measurement stage: the dynamics is still simulated in the full Hilbert space, and benefits from the exact analogy. 

\subsection{2$\times$2 plaquettes } \label{sec:hybrid:2x2}
\begin{figure}
\centering
  \includegraphics[width=0.65\columnwidth]{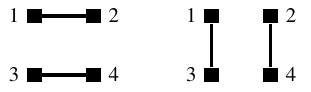}
  \caption{The model with $2\times 2$ plaquettes (each represented by a black square) can be quantum simulated by splitting the the interactions into the horizontal part (left) and the vertical part (right). This allows for an effective implementation of the full Hamiltonian via Trotterization.}
  \label{fig:2x2}
\end{figure}

Assuming that tunable capacitors are available, we now have to come up with feasible partitions of the Hamiltonian into parts that include only pairwise interactions.
For $2\times 2$ plaquettes we can split it in two parts: one with the horizontal interactions and the other with the vertical ones.
This means alternating between the two inverse capacitance matrices 
\begin{equation} \label{eqn:2x2_trotter_inv_mats}
    \parsq{C^{-1}_\text{hor}} \propto \begin{pmatrix}
        2 & 1 & 0 & 0 \\
        1 & 2 & 0 & 0 \\
        0 & 0 & 2 & 1 \\
        0 & 0 & 1 & 2
    \end{pmatrix}, 
    \hspace{10pt}
    \parsq{C^{-1}_\text{ver}} \propto \begin{pmatrix}
        2 & 0 & 1 & 0 \\
        0 & 2 & 0 & 1 \\
        1 & 0 & 2 & 0 \\
        0 & 1 & 0 & 2
    \end{pmatrix}, 
\end{equation}
where the four plaquettes/junctions are numbered according to \figref{fig:2x2}. 
Note that we have $2$ on the diagonal (and not $4$): the diagonal matrix elements have to be scaled by $1/2$ compared to the analog matrix \eqref{eqn:analog_inv_C}, such that after Trotterization the effective Hamiltonian will have the correct coefficient before $\op{n}_i^2$.

This can be implemented by making the coupling capacitors $C_{12}, C_{34}, C_{13}, C_{24}$ tunable, with some on-value $C_\text{on}$, and designing the (fixed) self capacitances to the same value $C_1 = C_2 = C_3 = C_4 = C_\text{on}$.
The capacitance matrix $\parsq{C_\text{hor}}$ is implemented when $C_{12}, C_{34}$ are on and $C_{13}, C_{24}$ are off, and $\parsq{C_\text{ver}}$ is implemented in the complementary case.
Since the $-E_J \cos{\op{\phi}_i}$ terms are always on, the effective Hamiltonian after Trotterization is
\begin{equation} \label{eqn:2x2_trotter_ham}
    \op{H}_\text{eff} = -2E_J \sum_{i=1}^4\cos{\op{\phi}_i} + \frac{1}{2} \frac{4e^2}{3 C_\text{on}}\paren{4\sum_{i=1}^4 \op{n}_i^2 + 2\sum_{\left< i,j\right>}\op{n}_i \op{n}_j}.
\end{equation}
Like in section \ref{sec:analog}, we can bring this Hamiltonian closer to the correct form \eqref{eqn:HU1_final} by tuning 
\begin{equation} \label{eqn:2x2_EJ}
    E_J = \frac{2e^2}{3C_\text{on}} \frac{1}{g^4}, 
\end{equation}
such that 
\begin{equation} \label{eqn:2x2_analogy}
       \op{H}_\text{eff} = \frac{4e^2}{3 C_\text{on}} \frac{1}{g^2} \op{H}_{\text{U(1)}}.
\end{equation}

Since it requires only 4 junctions (or 4 pairs of junctions - for a $g^2$-tunable experiment), and 4 tunable capacitors, the complexity of the circuit is quite modest, and we believe that an experimental implementation is not far off, paving the ground for a truncation-free QS for a non-trivial system. 
We acknowledge that the tunable capacitor design has not yet been tested experimentally, which makes the feasibility somewhat less certain; however the numerical analysis of \fullcite{Materise_2023_Tunable_capacitor} seems very thorough, and we believe that it is only a matter of time before a working example is provided. 

\subsection{1D chain of plaquettes} \label{sec:hybrid:1D}
\begin{figure}
\centering
  \includegraphics[width=0.95\columnwidth]{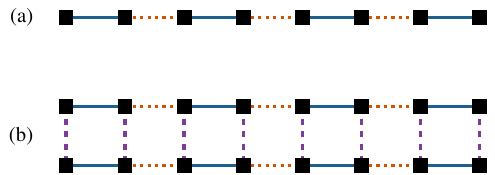}
  \caption{(a) The one dimensional chain of plaquettes (each represented by a black square) can be quantum simulated by splitting the interactions into the odd part (solid blue) and the even part (dotted orange). (b) For the dual rail of $2\times N$ plaquettes the interactions are split into three pairwise parts: horizontal odd (solid blue), horizontal even (dotted orange), and vertical (dashed purple).}
  \label{fig:1xN_2xN}
\end{figure}

For a chain of $N$ plaquettes, the natural way to split the Hamiltonian is again in two parts, namely the odd interactions and the even interactions, as indicated in \figref{fig:1xN_2xN}(a).
This works much the same as the $2\times 2$ case, but with a small complication at the boundaries of the chain.
Assuming (for concreteness) that $N$ is even, the required inverse capacitance matrices are
\begin{equation} \label{eqn:1D_trotter_inv_mats}
    \parsq{C^{-1}_\text{odd}} \propto \begin{pmatrix}
        2 & 1 &   &  &  &  &\\
        1 & 2 &   &  &  &  &\\
          &   & 2  & 1 &  &  &  \\
          &   & 1 & 2 &   &  &   \\
          &   &   &    & \ddots  &   \\
          &   &   &   &   & 2 &  1 \\        
          &   &   &   &  &  1 & 2
        
    \end{pmatrix}, 
    \hspace{3pt}
    \parsq{C^{-1}_\text{even}} \propto \begin{pmatrix}
        2 &   &   &  &  &  &\\
          & 2 & 1 &  &  &  &\\
          & 1 & 2  &  &  &  &  \\
          &   &   & 2 & 1 &  &   \\
          &   &   & 1  & 2  &   \\
          &   &   &   &   & \ddots &   \\        
          &   &   &   &  &   & 2
    \end{pmatrix}. 
\end{equation}
While in the bulk each JJ is always coupled to exactly one other junction, the two boundary junctions are isolated in the even part, and coupled (to their neighbors) during the odd part.  
This means that we are not going to be able to use a fixed self-capacitance value for the two boundary junctions.

Specifically, by inverting \fulleqref{eqn:1D_trotter_inv_mats} we realize that in the bulk we can implement as before, with fixed self capacitances that equal the on-value of the tunable coupling capacitors
\begin{equation} \label{eqn:1D_cap_values_bulk}
   C_i = C^\text{on}_{i,i+1} \equiv C_\text{on} \hspace{30pt}  \text{for $1<i<N$}.
\end{equation}
However, at the boundaries we have to alternate between two different values when implementing the two Trotterization parts: 
\begin{equation} \label{eqn:1D_cap_values_boundaries}
   C_1 = C_N = \begin{cases}
       C_\text{on} & \text{for the odd part}\\
       3 C_\text{on}/2 & \text{for the even part}
   \end{cases}
\end{equation}
for even $N$ (and the adjustment for odd $N$ is straightforward).
In practice it means that the boundary junctions have to be shunted to the ground via  tunable capacitors, rather than fixed ones.
Tuning the Josephson energy according to \fulleqref{eqn:2x2_EJ} and implementing this Trotterization procedure with the alteration of the self capacitances at the boundaries, the effective Hamiltonian obeys again \fulleqref{eqn:2x2_analogy}, and equals the model Hamiltonian up to a prefactor that depends on $g^2$.

\subsection{Dual rail of plaquettes} \label{sec:hybrid:semi2D}
Our next proposal is for a quasi two dimensional dual rail of $2\times N$ plaquettes.
Because of the restriction to have only pairwise interactions in each Trotter part, here we have to split the Hamiltonian into three parts: odd horizontal interactions, even horizontal interactions, and vertical interactions, as showed in \figref{fig:1xN_2xN}(b).
The three required inverse capacitance matrices are constructed from the following single-pair block
\begin{equation} \label{eqn:ladder_Cinv_pair}
    \parsq{C^{-1}_\text{pair}}\propto \begin{pmatrix}
        {4}/{3} & 1 \\
        1 & {4}/{3}
    \end{pmatrix},
\end{equation}
acting on the relevant subset of plaquette-pairs (horizontal-odd, horizontal-even or vertical).
The value $4/3$ is a result of splitting the Hamiltonian into three parts, which means that the diagonal value has to be scaled by $1/3$ relative to \fulleqref{eqn:analog_inv_C}.
As before, this can be implemented with tunable coupling capacitors with some on-value $C_\text{on}$ between neighbouring junctions on the (dual) lattice, but with the self capacitances designed to $C_\text{on}/3$ in the bulk. 
In the left and right boundaries, tunable self capacitances are needed like in section \ref{sec:hybrid:1D}, alternating between the value of $C_\text{on}/3$ when they participate in an interaction (during the horizontal-odd and vertical parts) and $7C_\text{on}/12$ when they do not (during the horizontal-even part).
The Josephson energy has to be designed or tuned to $E_J = 12e^2/\paren{7C_\text{on}g^4}$ such that
\begin{equation} \label{eqb:H_eff_ladder}
 \op{H}_\text{eff} = \frac{36e^2}{7C_\text{on}} \frac{1}{g^2} \op{H}_\text{U(1)}.   
\end{equation}

This does not work for more than two rails ($3\times N$ plaquettes or more) because in that case each plaquette participates in 4 interactions, and consequently we have to split the Hamiltonian in 4 Trotter parts. 
This means that the inverse capacitance matrices in each part should be constructed from the pairwise-blocks that obey
\begin{equation} \label{eqn:Cinv_pair_3xN}
    \parsq{C^{-1}_\text{pair}} \propto \begin{pmatrix}
        1 & 1\\
        1 & 1 
    \end{pmatrix},
\end{equation}
where as before the value on the diagonal is the value from \fulleqref{eqn:analog_inv_C}, scaled down by the number of parts.
This matrix is non-invertible, and therefore no physical circuit implements it. 
We can always split the Hamiltonian into more parts such that the required matrix is not singular, but in that case it would still be non-physical, as we explain in the following subsection.

\subsection{Explaining the numerical prefactors} \label{sec:prefactors}
In the three hybrid proposals presented above, we came across many numerical prefactors (e.g. in the required $E_J$ or self capacitance values).
Here we show the more general procedure to derive them, which will also explain why splitting the Hamiltonian to more than 3 parts is not possible.

By assumption, each junction is either coupled to one other junction (participates in a pairwise interaction), or to none at all.
Bulk junctions are always of the first kind, and for boundary junctions it depends on the specific partition of the Hamiltonian into Trotter parts.
This means that the capacitance matrix is constructed out of $2\times 2$ blocks, which we denote as $\parsq{C_\text{pair}}$; and $1\times 1$ blocks, which we denote as $\parsq{C_\text{single}}$.
As before we denote the on-value of the coupling capacitors as $C_\text{on}$, and it is clear from section \ref{sec:hybrid:1D} that the self capacitances should depend on whether or not the junction is part of pair (in the current Trotter part):
\begin{equation} \label{eqn:definition_Cs}
    C_i = \begin{cases}
        \alpha C_\text{on} & \text{if $i$ is part of pair}\\
        \beta C_\text{on} & \text{if $i$ is not part of pair},\\
    \end{cases}
\end{equation}
where $\alpha$ and $\beta$ are two non-negative dimensionless values. 

The normalized capacitance matrix $\parsq{c}\equiv \parsq{C}/C_\text{on}$ is constructed out of the following blocks:

\begin{align}
    &\parsq{c_\text{pair}} =  \begin{pmatrix}
    1+\alpha & -1 \label{eqn:c_pair} \\
    -1 & 1+\alpha
        \end{pmatrix}\\
    &\parsq{c_\text{single}} = \beta, \label{eqn:c_single} 
\end{align}
and therefore the inverse capacitance matrix is constructed from their inverses:
\begin{align}
    &\parsq{c^{-1}_\text{pair}} = \frac{1}{\abs{c_\text{pair}}} \begin{pmatrix}
    1+\alpha & 1 \\
    1 & 1+\alpha
        \end{pmatrix} \label{eqn:c_pair_inv}\\
    &\parsq{c^{-1}_\text{single}} = \frac{1}{\beta} = \frac{1}{\abs{c_\text{pair}}}\frac{\abs{c_\text{pair}}}{\beta}.
\end{align}
If the Hamiltonian is divided into $p$ parts, then by the scaling condition we need 
\begin{align}
        \alpha &= \frac{4}{p} - 1 \label{eqn:alpha}\\
        \beta &= \frac{p \abs{c_\text{pair}}}{4} = \frac{4}{p} - \frac{p}{4}. \label{eqn:beta}
\end{align}
Additionally, since the magnetic part is always on, the effective Hamiltonian has $pE_J$ in front of the magnetic terms. This means that to get the correct ratio to the electric terms we need to design or tune
\begin{equation}  \label{eqn:E_J_general_p}
    E_J = \frac{4e^2}{p \abs{c_\text{pair}}C_\text{on}} \frac{1}{g^4} = \frac{1}{\beta}\frac{e^2}{C_\text{on}}\frac{1}{g^4}
\end{equation}
such that the effective Hamiltonian obeys
\begin{equation} \label{eqn:H_eff_general_p}
    \op{H}_\text{eff} = \frac{4e^2}{\abs{c_\text{pair}}C_\text{on}} \frac{1}{g^2} \op{H}_\text{U(1)}.
\end{equation}
\fulleqref{eqn:alpha} - \eqref{eqn:H_eff_general_p} are consistent with the values given in sections \ref{sec:analog} and \ref{sec:hybrid:2x2}-\ref{sec:hybrid:semi2D} for $p=1,2,3$.

Note that if $p=4$ the matrices are non-invertible, and for $p>4$ they are not physical (implying negative self-capacitance $\alpha<0$). 
For this reason the hybrid method as described here only works for $p=2,3$, and cannot be extended for the fully-2D model which requires $p\ge 4$ to cover all interactions while using only pairwise Trotter parts. 
In section \ref{sec:2D} we suggest an alternative approach that solves this problem, but comes with more difficult experimental requirements. 

\section{Imperfect tunable capacitors} \label{sec:errors}
Consider the $2 \times 2$ plaquettes proposal from section \ref{sec:hybrid:2x2}, but this time assume that the tunable coupling capacitors have a large-but-finite on/off ratio $1/\eta$, such that $\eta$ is a small parameter. 
In this case, in the first (horizontal) Trotter part we are implementing the capacitance matrix
\begin{equation} \label{eqn:error2x2_cap_1}
    \parsq{C_\text{hor}} ={C_\text{on}} \begin{pmatrix}
        2+\eta & -1 & -\eta & 0 \\
        -1 & 2+\eta & 0 & -\eta \\
        -\eta & 0 & 2+\eta & -1 \\
        0 & -\eta & -1 & 2+\eta
    \end{pmatrix}.
\end{equation}
As a first easy correction to bring the this closer to the ideal form, we can change the value of the self capacitors from $C_i = C_\text{on}$ to 
\begin{equation} \label{eqn:error_Ci_step1}
    C_i = C_\text{on}\paren{1-\eta},
\end{equation}
which results in
\begin{equation} \label{eqn:error2x2_cap_2}
    \parsq{C_\text{hor}} ={C_\text{on}} \begin{pmatrix}
        2 & -1 & -\eta & 0 \\
        -1 & 2 & 0 & -\eta \\
        -\eta & 0 & 2 & -1 \\
        0 & -\eta & -1 & 2
    \end{pmatrix}.
\end{equation}
Inverting \fulleqref{eqn:error2x2_cap_2} and keeping up to first order in $\eta$, we find
\begin{equation} \label{eqn:error2x2_cap_inv}
    \parsq{C^{-1}_\text{hor}} =\frac{1}{3C_\text{on}} \parsq{\begin{pmatrix}
        2 & 1 & 0 & 0 \\
        1 & 2 & 0 & 0 \\
        0 & 0 & 2 & 1 \\
        0 & 0 & 1 & 2
    \end{pmatrix}
    + \frac{\eta}{3} \begin{pmatrix}
        0 & 0 & 5 & 4 \\
        0 & 0 & 4 & 5 \\
        5 & 4 & 0 & 0 \\
        4 & 5 & 0 & 0
    \end{pmatrix}
     + O\paren{\eta^2}},
\end{equation}
 which is the implemented inverse-capacitance matrix written as a sum of the ideal matrix and the leading order deviation, and a similar result can be obtained for $\parsq{C^{-1}_\text{ver}}$.
 
 The deviation introduces unwanted interactions, and it is useful to separate them into two different types: (1) the interactions that result from the anti-diagonal of the deviation matrix (these are diagonal interactions on the $2\times 2$ lattice, see \figref{fig:2x2}), and (2) the other unwanted  interactions (these are vertical interactions on the $2\times 2$ lattice).
 The reasoning behind this separation is that while the interactions of type (1) are completely unwanted, the interactions of type (2) are only unwanted in the horizontal part of the Trotterization, but they are in fact exactly the vertical interactions that we do want to implement in the second part.
 This means that the type (2) errors can be corrected by further engineering the ratio $C_i/C_\text{on}$ such that the combined interaction strength from both parts will give the desired value. 
 Specifically, one can write the capacitance matrix for an arbitrary $\alpha \equiv C_i/C_\text{on}$ (for all $i=1,2,3,4$), and calculate the inverse to first order in $\eta$.
 Then require that the ratio between the diagonal elements and the sum of the wanted elements and the type (2) elements will be $2:1$, which for the $2 \times 2$ plaquettes system reduces to (see the Appendix)
 \begin{equation} \label{eqn:error2x2_condition}
     1 - \alpha + \frac{3\alpha^2 + 2\alpha + 2}{ \alpha^2 + 2\alpha}\eta=0.
 \end{equation}

 In total, if $\alpha = 1-\eta$ (chosen to obey \fulleqref{eqn:error_Ci_step1}), and $E_J$ is tuned according to \fulleqref{eqn:2x2_EJ}, then the effective Trotterized Hamiltonian obeys
 \begin{equation} \label{eqn:error_2x2_Ham}
     \op{H}_\text{eff}= \frac{4e^2}{3C_\text{on}}\frac{1}{g^2}\paren{\op{H}_\text{U(1)} + g^2\parsq{\frac{2  \eta}{3} \paren{8\op{H}^{\paren{1}}_\text{err} + 5 \op{H}^{\paren{2}}_\text{err} } + \mathcal{O}\paren{ \eta^2}}},
 \end{equation}
 where
 \begin{equation} \label{eqn:2x2_H_err_1}
     \op{H}^{\paren{1}}_\text{err} = {\op{n}_1 \op{n}_4 + \op{n}_2 \op{n}_3} 
 \end{equation}
 is the unwanted interaction of type (1), and
 \begin{equation} \label{eqn:2x2_H_err_2}
     \op{H}^{\paren{2}}_\text{err} =\sum_{\left<i,j\right>}\op{n}_i \op{n}_j 
 \end{equation}
 is the unwanted interaction of type (2). 
 On the other hand, choosing $\alpha$ to obey \eqref{eqn:error2x2_condition}, we get rid of $\op{H}^{\paren{2}}_\text{err}$, but then a different $E_J$ value is required, and the numerical prefactors in \fulleqref{eqn:error_2x2_Ham} also change (for the details see the Appendix).

\begin{figure}[h]
    \centering
    \begin{minipage}{0.48\textwidth}
        \centering
        \includegraphics[width=\textwidth]{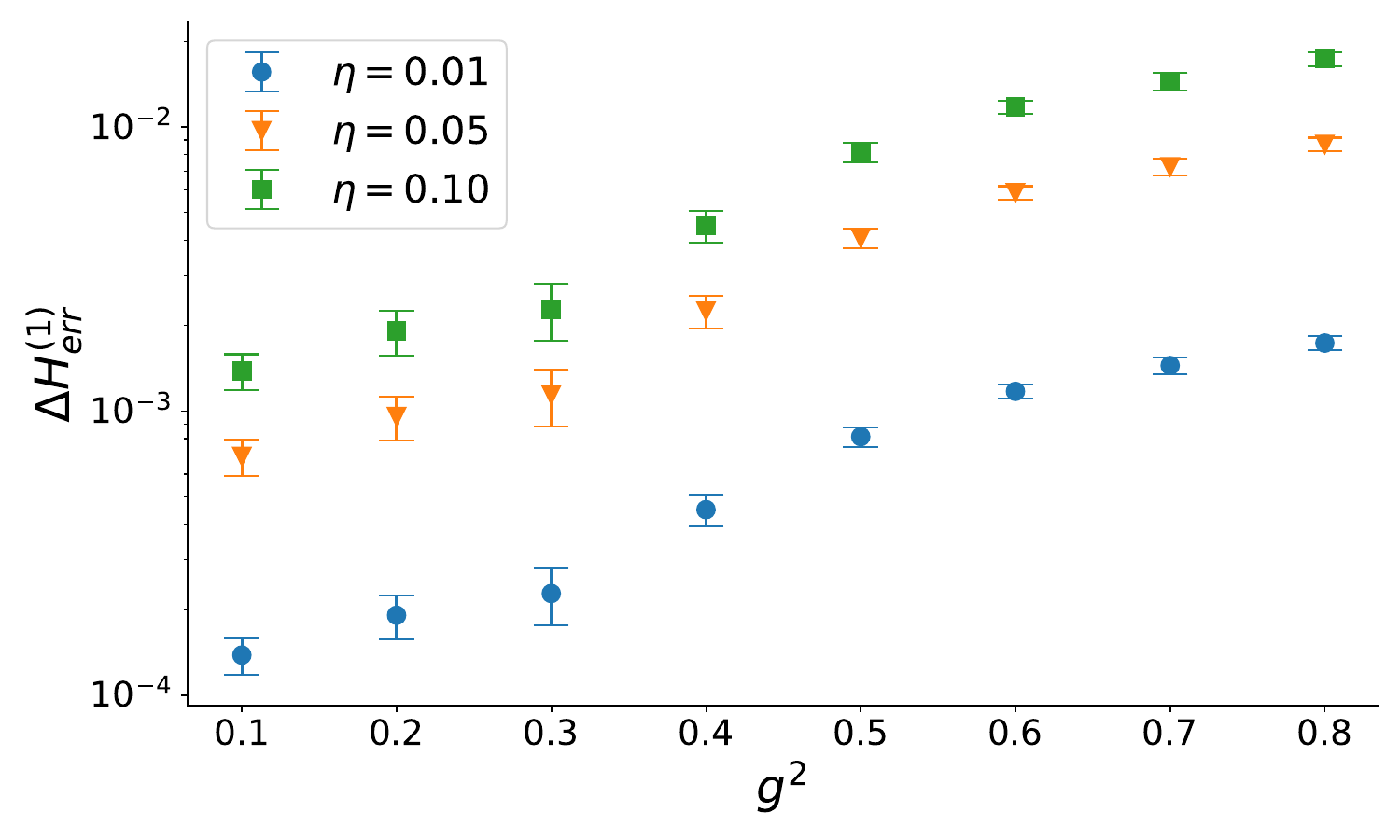}
    \end{minipage}
    \hspace{0.2cm}
    \begin{minipage}{0.48\textwidth}
        \centering
        \includegraphics[width=\textwidth]{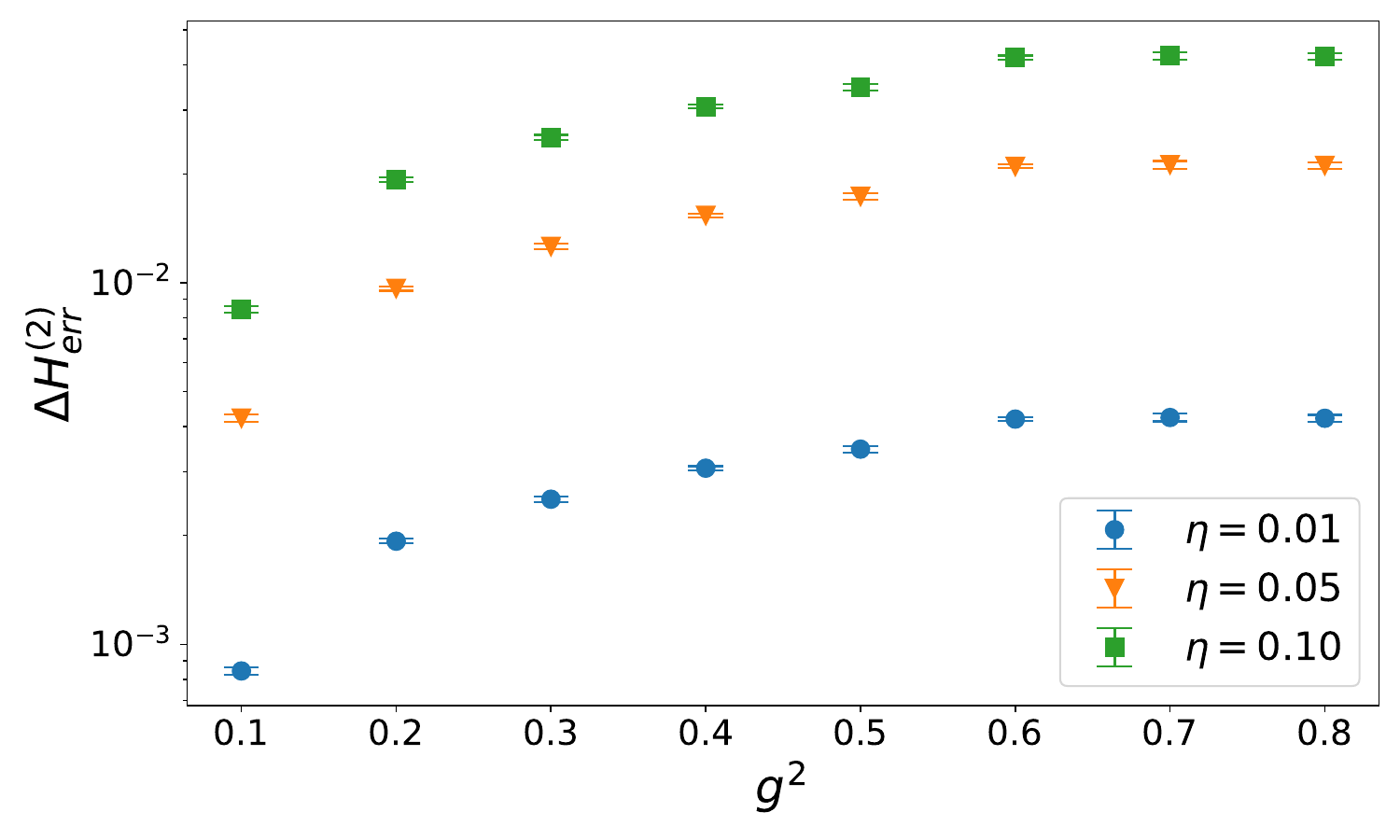}
    \end{minipage}
    \caption{Numerical estimates of the errors due to finite on-off ratios in the tunable capacitors on a $2 \times 2$ plaquettes lattice. We plot the absolute value of the error normalized by the number of plaquettes, $\Delta H_{\text{err}} = \abs{\braket{\op{H}_{\text{err}}}}/N_{\text{plaq}}$, where $\op{H}_\text{err}$ is the error Hamiltonian, as derived from \fulleqref{eqn:error_2x2_Ham}. Hence, we can compare these values to the ground state energy per plaquette which is slightly below one in the coupling regime we are considering here. In the upper plot we show the result for errors of type (1) and in the lower plot the errors of type (2). We consider different on/off ratios $1/\eta$ and see that for on/off ratios that are achievable under realistic assumptions (\fullcite {Materise_2023_Tunable_capacitor} predicts $\eta \approx 0.025$), the error is relatively moderate.}
    \label{fig:error_estimate}
\end{figure}

Since it seems hard to obtain analytical estimates for the  the errors due to the charge operators $\hat{n}_i$ being unbounded, we employ a recently developed variational Monte Carlo method~\cite{bender_variational_2023} to get numerical estimates for these errors. We minimize a variational ground state energy with respect to the exact Hamiltonian over a region of coupling constants $g^2$ and compute the magnitude of type (1) and type (2) errors in the resulting state.
This provides an estimate to the first order correction to the ground state energy.
The result is shown in \figref{fig:error_estimate}, demonstrating that for reasonable $\eta$ values (\fullcite{Materise_2023_Tunable_capacitor} predicts $\eta \approx 0.025$ with realistic parameters) the magnitude of type (1) and type (2) errors becomes negligible compared to the ground state energy per plaquette which is $\mathcal{O}(1)$. 
Furthermore, the larger of the two error types (by about an order of magnitude, for small $g^2$) is the one we can correct by choosing a more optimal value for the fixed capacitors.

 We obtain similar results when carrying out the same error analysis for the 1D chain of $N$ plaquettes. 
 Here we can use again \fulleqref{eqn:error_Ci_step1} as a first easy correction, but only for the bulk junctions $1<i<N$. 
 On the boundary junctions we have to use tunable self-capacitors (see sections \ref{sec:hybrid:1D} and \ref{sec:prefactors}), and alternate their values during the experiment depending on whether they participate in a pairwise interaction in the current Trotter part or not.
 If $N$ is even, what we need is
 \begin{equation} \label{eqn:error_1D_coundary_Ci}
     C_1 = C_N = \begin{cases}
         C_\text{on} & \text{for the odd part}  \\
         C_\text{on}\paren{\frac{3}{2}-\eta} & \text{for the even part},
     \end{cases}
 \end{equation}
and the adjustment for odd $N$ is straightforward.
To first order in $\eta$, the error is local in the loop variables $\op{n}_i$, and the effective Hamiltonian is
\begin{equation} \label{eqn:error_1d_H_eff}
    \op{H}_\text{eff} = \frac{4e^2}{3C_\text{on}} \frac{1}{g^2} \parsq{\op{H}_\text{U(1)} + g^2\paren{\frac{2\eta}{3}  \op{H}_\text{err}  + \mathcal{O}\paren{\eta^2}}},
\end{equation}
where
\begin{equation}
     \op{H}_\text{err}  =4\sum_{i=1}^{N-1} \op{n}_i \op{n}_{i+1} +  4\sum_{i=1}^{N-2} \op{n}_i \op{n}_{i+2}  + \sum_{i=1}^{N-3} \op{n}_i \op{n}_{i+3}.
\end{equation}
The first sum (over nearest-neighbours) is an error of type (2), meaning that it is an interaction that appears also in the correct Hamiltonian. 
Therefore it can be corrected via the protocol that is described in the Appendix. 

A similar calculation can be performed for the dual rail ladder with $2\times N$ plaquettes, with the same qualitative results.
Namely, the finite on/off ratio introduces errors that are local to first order in $\eta$, with maximal interaction range of 3 plaquettes.
The nearest-neighbours errors can in principle be fixed out, but the other terms remain and for the QS to work we need them to be small compared to $\op{H}_\text{U(1)} $, meaning (for all three proposals) 
\begin{equation} \label{ean:error_eta_condition}
    \eta\ll 1 , \hspace{10pt} \eta\ll \frac{1}{g^4}.
\end{equation}
Since QS is mostly relevant for the $g^2<1$ regime (see section \ref{sec:background:U1LGT}), it is enough to require $\eta \ll 1$.
To improve precision while working with a moderately small $\eta$, one can use a method similar to zero noise extrapolation (ZNE).
By repeating the experiment a few times while increasing the off-values of the tunable capacitors we can effectively implement different values of $\eta$, and then extrapolate the resulting observables to $\eta=0$. Such methods have been shown to significantly improve the precision of evaluated observables in noisy experiments \cite{Giurgica-Tiron_ZNE_2020}.

\section{Progress towards the fully 2D model} \label{sec:2D}
In section \ref{sec:hybrid} we could not generalize the methods to the fully-2D model because the required inverse capacitance matrix is constructed from $2\times 2$ blocks that obey \eqref{eqn:Cinv_pair_3xN}, which is non-invertible.
Here we propose using an approximate matrix which is invertible. 
Again we use tunable coupling capacitors with on value $C_\text{on}$, and assume that the self capacitances obey $C_i =  \alpha C_\text{on} \equiv C_J$ for all the junctions $i$ in the bulk.
As before, we will have to use tunable capacitors in the boundaries, with alternating values that have to be calculated.
However, for the purpose of of estimating the feasibility of this method we will focus only on the bulk. 

The (normalized) pair capacitance matrix is given by \eqref{eqn:c_pair}, and the inverse by \eqref{eqn:c_pair_inv}.
Assuming $\alpha \ll 1$ (designing the self capacitance $C_J$ to be as small as possible), the inverse capacitance matrix is 
\begin{equation} \label{eqn:2D_Cinv_pair}
    \parsq{C^{-1}_\text{pair}} = \frac{1}{2C_J}\parsq{ \begin{pmatrix}
        1 & 1 \\
        1 & 1
    \end{pmatrix}
     + \frac{\alpha}{2}\begin{pmatrix}
        1 & -1 \\
        -1 & 1
    \end{pmatrix} + \mathcal{O}\paren{\alpha^2}}
\end{equation}
which means that the pair Hamiltonian is 
\begin{equation} \label{eqn:2D_pair_ham}
\begin{aligned}
    \op{H}_{ij} = &-E_J\paren{\cos{\op{\phi}_i} + \cos{\op{\phi}_j}} \\
    &+ \frac{e^2}{C_J} \parsq{\paren{1+\frac{\alpha}{2}}\paren{\op{n}^2_i + \op{n}^2_j} + 2\paren{1-\frac{\alpha}{2} }\op{n}_i\op{n}_j + \mathcal{O}\paren{\alpha^2} }.
\end{aligned}
\end{equation}
Since in this method the model Hamiltonian is divided in $p=4$ parts, after Trotterization we will have implemented the effective Hamiltonian
\begin{equation} \label{eqn:2D_eff_ham}
\begin{aligned}
    \op{H}_\text{eff} = &-4E_J\sum_i \cos{\op{\phi}_i} \\
    &+ \frac{e^2}{C_J} \parsq{4\paren{1+\frac{\alpha}{2}}\sum_i \op{n}^2_i + 2\paren{1-\frac{\alpha}{2} } \sum_{\left< i,j\right>}\op{n}_i\op{n}_j  + \mathcal{O}\paren{\alpha^2}},
\end{aligned}
\end{equation}
Which has the correct form up to a local error that include $\op{n}_i^2$ terms and nearest-neighbors $\op{n}_i\op{n}_j$ terms. 
By tuning $E_J$, we have the freedom to choose which type of error will be more dominant in the leading order.
For example, to have only $\op{n}_i^2$ errors we choose 
\begin{equation} \label{eqn:EJ_2D1}
    E_J = \paren{1-\frac{\alpha}{2}} \frac{e^2}{2C_J}\frac{1}{g^4},
\end{equation}
such that 
\begin{equation} \label{eqn:2D_final_Heff1}
    \op{H}_\text{eff} = \paren{1-\frac{\alpha}{2}} \frac{2e^2}{C_J}\frac{1}{g^2}\parsq{\op{H}_\text{U(1)} + g^2\paren{2\alpha \sum_i \op{n}_i^2 + \mathcal{O}\paren{\alpha^2}}}.
\end{equation}
Alternatively we can choose to have only the nearest neighbors error, tuning
\begin{equation} \label{eqn:EJ_2D2}
    E_J = \paren{1+\frac{\alpha}{2}} \frac{e^2}{2C_J}\frac{1}{g^4},
\end{equation}
such that
\begin{equation} \label{eqn:2D_final_Heff2}
    \op{H}_\text{eff} = \paren{1+\frac{\alpha}{2}} \frac{2e^2}{C_J}\frac{1}{g^2}\parsq{\op{H}_\text{U(1)} - g^2\paren{\alpha \sum_{\left<i,j\right>} \op{n}_i\op{n}_j + \mathcal{O}\paren{\alpha^2}}};
\end{equation}
and it is also possible to choose intermediate $E_J$ values such that both errors exist, with some chosen ratio of amplitudes.
Similar to the on/off-ratio error from section \ref{sec:errors}, for the error to be small we need both $\alpha\ll 1$ and $\alpha\ll 1/g^4$, but the first condition is enough if $g^2<1$ (which is the more interesting regime, see section \ref{sec:background:U1LGT}).
Also here ZNE-inspired methods can help mitigate the errors even if $\alpha$ cannot be made vanishingly small. However, since $\alpha$ parametrizes a fixed capacitance value, in this case the ZNE has to be implemented by fabricating a few different versions of the circuit.

The smallest reasonable self capacitance is of the order of $C_J\approx 1 \text{fF}$ \cite{Shimazu_effects_1997, Watanabe_Coulomb_2001, Watanabe_quantum_2003}, while for the coupling capacitors we can comfortably use $\sim 0.1-1 \text{pF}$, to get $\alpha\ll1$ as needed. These values imply (from \eqref{eqn:EJ_2D1}, \eqref{eqn:EJ_2D2} and \eqref{eqn:I_c})  that the critical current of the junctions should be around $I_c\approx  40 \text{nA}/g^4 $.
While $40 \text{nA}$ is a comfortable value, for small $g^2$ the required $I_c$ can quickly get quite far from standard.
Depending on the desired $g^2$ values, we might need to make $\alpha$ larger and sacrifice accuracy in order to work with manageable circuit parameter values.

Another tradeoff that will have to be considered when designing the experiment has to do with the energy scale of the Hamiltonian.
Specifically, since $ 10 \text{GHz}$ is about as fast as standard rf-electronics operate,  if $2e^2/\paren{h C_J g^2} > 10 \text{GHz}$, turning the tunable capacitors on and off can no longer be assumed to be immediate.
This might compromise the Trotter approximation, and thus depending on the desired $g^2$ this can also require a larger $\alpha$.
However it does seem that pushing either $I_c$ or $C_\text{on}$ slightly beyond standard values can result in a feasible working point. 
For example, for $g^2=0.2$ if we choose $C_J\approx 100 \text{fF}$ and $C_\text{on}$ to be a few pF (such that $\alpha$ is a few percents), then the required $I_c\approx 10 \text{nA}$ and the characteristic frequency is in the GHz range.

The most critical weakness of this method is revealed when considering the finite on/off ratio of the coupling capacitors. In a similar procedure to the one described in section \ref{sec:errors}, we computed the deviations from the ideal Hamiltonian to first order in the two small parameters $\alpha$ and $\eta$.
We find again that the error is local in the loop variables, but critically, it includes some terms that scale as $\eta / \alpha$. 
This means that for the approximation to be valid we need $\eta\ll \alpha \ll 1$, which is going to be difficult with the tunable capacitors design of \fullcite{Materise_2023_Tunable_capacitor}, that predicts (numerically) $\eta\approx 0.025$.
A significant improvement (one order of magnitude or more) in the tunable capacitors technology is required to make this proposal feasible. 

\section{Summary and discussion} \label{sec:discussion_and_summary}
To conclude, in this work we propose to take advantage of the exact analogy at the level of the local Hilbert space between an array of JJs and a pure-gauge U(1) LGT, and to use it for QS. This method provides an opportunity for utilizing superconducting circuits as an analog platform (rather then digitally, with superconducting qubits), and could potentially be used to probe lattice QED at large system sizes and without truncating the Hilbert space.

Using the dual formulation of the model, we showed that an exact analogy can be established also at the level of the Hamiltonian for two plaquettes.
For larger systems we propose a hybrid analog-digital approach in which the full Hamiltonian is implemented effectively via a Trotter decomposition into pairwise parts that are implemented analogically. 
Tunable coupling capacitors are required for this to work, which is the main experimental/technological challenge that has to be solved before our proposal can be implemented.
In theory this issue is already solved by \fullcite{Materise_2023_Tunable_capacitor}, but an experimental demonstration of the tunable capacitor design has yet to be reported. 
Nevertheless this design is already used in a few theoretical proposals for quantum devices (e.g. \cite{Vilkelis_2024_fermionic}), and we are quite optimistic about it being implemented soon.
Since our proposal is based on the explicitly gauge-invariant dual formalism, any experimental error in the QS (such as the errors that are analyzed in section \ref{sec:errors}) would not cause a gauge violation.
This is important because it means that regardless of the severity of  the experimental errors, we can be absolutely certain that we implement a LGT (but maybe not exactly the one we wanted, if the errors are not controlled).

From an experimental perspective, the immediate next step would be to implement the analog two-plaquettes proposal which does not require any new components, and can be used to benchmark the method and to compare against different truncation schemes. 
Such a simulation would already be interesting since analytical solutions only exist for one plaquette and classically simulating the untruncated, infinite-dimensional Hilbert space becomes already difficult for a few plaquettes.
In parallel, building a proof-of-concept demonstration of the tunable capacitor design would be the first step towards implementing the more interesting hybrid proposals.
Beyond that, it can be useful to  develop a measurement protocol for non-local observables (Wilson loops). In principle this can be done with local operations, as shown in \fullcite{zohar_local_2020}; 
but it could be more efficient with a specifically designed physical global readout similar to \fullcite{Yang_2020_probing}, in which a single readout resonator is coupled to multiple transmon qubits.

From a theoretical perspective we see two possible directions that can be further investigated: the first of which is to consider a theory with matter.
The dual formulation has been already generalized to include fermionic matter \cite{bender_gauge_2020}, which in LGTs can be represented by superconducting qubits \cite{zohar_eliminating_2018,Pardo_2023_resource}.
Therefore it is plausible that one could design a circuit similar to our CCJAs, with additional transmon qubits to encode the matter. 
Another interesting direction is to investigate the relation between our analogy and the duality of \fullcite{HOSOTANI1977_compact}. 
As we speculated in section \ref{sec:intro}, the duality between a continuum QED and an extended-element (continuous) description of a JJ might be fundamentally related to the analogy between lattice QED and a lumped-element model for a JJ. 
Understanding this relation can potentially provide insight into the continuum limit of lattice theories in general.

\section*{Acknowledgments}
We would like to thank J. Ignacio Cirac for many fruitful discussions.
E.Z. acknowledges  the support of the Israel Science Foundation (grant No. 523/20). N.K. acknowledges support of EU project OpenSuperQPlus100.

\section*{Appendix: Correcting the on/off-ratio errors of type (2) } \label{app_a}
Here we provide further details on performing the partial correction of the on/off-ratio errors suggested in section \ref{sec:errors}. We show it here for the $2\times2$ plaquettes proposal, but this procedure can be readily adapted for the other proposals as well.
Writing the implemented capacitance matrix for a general $\alpha\equiv C_i/C_\text{on}$:
\begin{equation}
        \parsq{C_\text{hor}} ={C_\text{on}} \begin{pmatrix}
        1+\alpha +\eta & -1 & -\eta & 0 \\
        -1 & 1+\alpha +\eta & 0 & -\eta \\
        -\eta & 0 & 1+\alpha +\eta& -1 \\
        0 & -\eta & -1 & 1+\alpha +\eta
    \end{pmatrix},
\end{equation}
with the analogous expression for $\parsq{C_\text{ver}}$. Inverting and keeping terms up to first order in $\eta$, we find that 
\begin{equation}
     \parsq{C^{-1}_\text{hor}} \approx \frac{1}{ 3 C_\text{on}} \begin{pmatrix}
        d & w & u & a \\
        w & d & a & u \\
        u & a & d & w \\
        a & u & w & d
    \end{pmatrix}
    \hspace{10pt}
     \parsq{C^{-1}_\text{ver}} \approx \frac{1}{ 3 C_\text{on}} \begin{pmatrix}
        d & u & w & a \\
        u & d & a & w \\
        w & a & d & u \\
        a & w & u & d
    \end{pmatrix},
\end{equation}
where
\begin{align}
             d &= \frac{3\paren{\alpha + 1}} 
             {\alpha\paren{\alpha+2}} - \frac{3\paren{\alpha^2 + 2\alpha +2}}{\alpha^2\paren{\alpha+2}^2} \eta \label{eqn:d} \\ 
             w &=  \frac{3}{\alpha\paren{\alpha+2}} - \frac{6\paren{\alpha+1}}{\alpha^2\paren{\alpha+2}^2}\eta \\
             u &= \frac{3\paren{\alpha^2 + 2\alpha +2}}{\alpha^2\paren{\alpha+2}^2} \eta \label{eqn:u} \\
             a & = \frac{6\paren{\alpha+1}}{\alpha^2\paren{\alpha+2}^2}\eta.
\end{align}
To have the correct ratio between the nearest-neighbours $\op{n}_i \op{n}_j$ interactions and the $\op{n}_i^2$ terms in the effective Hamiltonian, we have to require
\begin{equation}
    d = 2\paren{w+u},
\end{equation}
which leads to \fulleqref{eqn:error2x2_condition} after substituting \eqref{eqn:d} -\eqref{eqn:u}.

If $\alpha$ is chosen to obey this condition, the effective Hamiltonian is
\begin{equation}
\begin{aligned}
    \op{H}_\text{eff} = &-2E_J \sum_{i=1}^4\cos{\op{\phi}_i}\\
    &+ \frac{w+u}{2} \frac{4e^2}{3 C_\text{on}}\paren{4\sum_{i=1}^4 \op{n}_i^2 + 2\sum_{\left< i,j\right>}\op{n}_i \op{n}_j + 4a\op{H}^{\paren{1}}_\text{err} + \mathcal{O}\paren{\eta^2}}.
\end{aligned}
\end{equation}
To get the U(1) Hamiltonian from this we need to tune
\begin{equation}
    E_J = \frac{2e^2 \paren{w+u}}{3C_\text{on}} \frac{1}{g^4},
\end{equation}
which results in 
\begin{equation}
    \op{H}_\text{eff} = \frac{4e^2 \paren{w+u}}{3C_\text{on}} \frac{1}{g^2}\parsq{\op{H}_\text{U(1)} + g^2\paren{ a \op{H}^{\paren{1}}_\text{err} + \mathcal{O}\paren{\eta^2} }},
\end{equation}
with $\op{H}^{\paren{1}}_\text{err}$ from \fulleqref{eqn:2x2_H_err_1}, the remaining type (1) error.
\bibliography{bibl}

\end{document}